\documentstyle[prb,aps,psfig]{revtex}

\begin{document}
\twocolumn[\hsize\textwidth\columnwidth\hsize\csname %
@twocolumnfalse\endcsname
\draft
\preprint{ver. 2.2}
%
%
\title{Identification of the physical parameters of the 
paramagnetic phase of the one-dimensional Kondo lattice model done by 
introducting a nonmagnetic quantum state with rotating order parameters}
\author{Mohamed Azzouz}
\address{Department of Physics, University of Northern British Columbia, 
Prince George, British Columbia, Canada
V2N 3Z9}
\date{\today}
\maketitle
\begin{abstract}
The paramagnetic phase of the one-dimensional Kondo 
lattice model is investigated for electron densities 
below half-filling using a new mean-field approach.  
The physical parameters that govern this phase
are identified to be the spin-flip processes of both the 
localized and itinerant spins.
A new nonmagnetic-quantum state, where the local magnetization is a rotating vector 
with a nonzero average length, is proposed in order to describe this phase.
this state does not break SU(2) symmetry in agreement with Mermin-Wagner theorem.
The line boundary between this phase and the ferromagnetic phase is calculated
in the coupling-density phase diagram. Also, expressions 
are calculated for 
the velocities of the conduction electrons excitations, and
heat capacity and entropy versus temperature are analyzed. 
Good agreement with many of the available numerical data is achieved.

\end{abstract}

\narrowtext
%
\section{INTRODUCTION}
Strong correlations between electronic degrees of
freedom in several materials, like high-temperature superconductors, 
heavy-fermion materials, etc., 
are responsible, or are at least thought to be responsible,  
for the occurrence of various quantum phenomena such as quantum magnetism, 
superconductivity, non-conventional metallic phases, etc. 
Heavy-fermion materials may be modeled using the canonical Kondo-lattice
model for which intensive studies have been reported in the last years
about its one-dimensional version.
In this case, the zero-temperature phase diagram in terms of the 
conduction-electrons density $n$ below half-filling 
and the Kondo exchange
coupling constant $J_K$ between the localized and itinerant spins 
has been established mainly using numerical 
methods.\cite{tsunetsugu1}$^-$\cite{shibata}
For strong exchange couplings, the ground
state is an unsaturated  ferromagnet, whereas
for weak couplings, 
it is a paramagnetic phase.
At exactly half-filling, the phase is a 
spin-liquid insulator for any Kondo coupling, Fig.\ 1 
(Ref. 1 contains detailed 
discussion of this phase diagram).
The ferromagnetic and the spin-liquid phases are now well understood
because these are characterized by {\it order parameters} 
which are the magnetization for the ferromagnetic phase, and 
the spin and charge energy gaps for the
spin-liquid phase.\cite{tsunetsugu1} For the paramagnetic phase however, no 
{\it physical parameter} has been proposed so far in order to describe
it. One of the purposes of this
work is to identify such a parameter (or parameters).
Moreover, the question concerning the crossover from 
the high-temperarture regime
to the low-temperature region 
needs to be addressed. 
In the high-temperature regime,
the conduction electrons 
and localized spins behave as being almost independent due to the fact that thermal
fluctuations wash out any characteristic energy related to the 
Kondo exchange coupling.
In the low-temperature regime, correlation effects become 
strong enough to dominate the physical properties of the Kondo lattice model.
The paramagnetic phase
was proposed to be a Luttinger liquid at zero 
temperature.\cite{tsunetsugu,sikkema,shibata} If we were to 
assume that this true, then
a crossover would take place between the high-temperature
metallic and the low temperature Luttinger-liquid phases.

Our findings can be summarized as follows.
The paramagnetic phase is found to be
governed by the quantum dynamics (spin-flip processes) of both 
the localized and itinerant spins.
We pave the way to a new quantum state in which the local magnetization is finite
in the xy-plane if measured in a rotating reference 
frame, but vanishes once averaged
over the angle of rotation, hence ensuring that Mermin-Wagner 
theorem is not violated;
the magnetization along the z-axis being equal to zero. This is an example of
a rotating {\it order prameter} with a finite length but a phase angle 
assuming any value between 0 and $2\pi$. We should mention that
interpreting our findings in terms of the Luttinger liquid state is not
a simple matter. We are still investigating this question.
Note however that quantitative comparison between our results
and the Luttinger-liquid description results is found to be 
fairly good (see below). What remains to do 
is understanding: why should the Fermi liquid picture give rise to the Luttinger liquid
one at low temperature?

In section II, a full description of the present approach  is provided. 
The physical foundations upon which our theoretical calculations
are based, are explained. We identify the physical parameters as being 
the averages of spin flip operators of the localized and itinerant electrons. 
Section III is devoted to showing some results and their discussions. 
The temperature
dependence of the parameters, crossover temperature, phase-diagram
boundary between the ferromagnetic and paramagnetic phases, 
as well as some of thermodynamic functions are reported.
Comparison with numerical data turns out to be very satisfactory.
In section IV, conclusions are drawn about the validity of the present approach.
\section{Description of the approach}
\subsection{Physical foundations}
In this work, the numerical data available in the literature are considered as 
{\it numerical experiments} upon which the present approach is founded.
\begin{figure}
\centerline{\psfig{figure=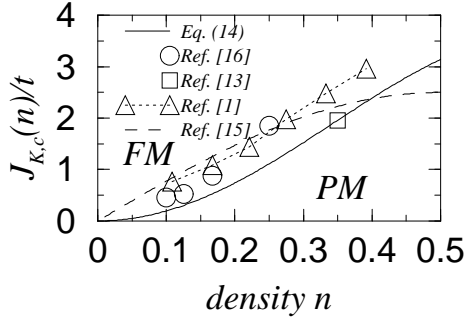,height=4.5cm,angle=0}}
\caption{The phase diagram obtained using Eq. 14 is shown
and compared to numerical results. FM, ferromagnetic phase. PM, 
paramagnetic phase.}
\label{MPQ}
\end{figure}
The Kondo lattice Hamiltonian in one dimension is given by:
\begin{equation}
H=-t\sum_{i,\sigma}(c_{i,\sigma}^{\dag}c_{i+1,\sigma} + {\rm H.c})
+ J_K\sum_{i}{\bf S}_i\cdot{\bf s}_i,
\label{hamiltonian0}
\end{equation}
where ${\bf S}_i$ and ${\bf s}_i$ are respectively the 
localized and itinerant spin operators.
$c^{\dag}_{i,\sigma}$ and $c_{i,\sigma}$ are the creation and annihilation 
operators at site $i$ of an electron with the z-component 
of the spin being $\sigma=\pm1/2$. $t$ is the hoping energy of 
conduction electrons. Because
neither the Kondo singlets formation nor magnetic order  
occurs in the paramagnetic phase for densities below half-filling, 
a result that is strongly 
pointed out by numerical calculations,\cite{tsunetsugu1}
it is justified to choose the following canal for decoupling the 
interacting term of 
(\ref{hamiltonian0}) where only the averages
$\langle s_i^-\rangle$ and $\langle S_i^-\rangle$ 
and their complex conjugates are taken into account.
These parameters represent spin-flip processes, and 
are therefore a good probe of the quantum dynamics of the system.
We suppose that these processes dominate over all other processes which are 
the Kondo screening and magnetic ordering. We will see that 
our results turn out to be consistent with this hypothesis.
It is worth noting here that in order to avoid breaking SU(2) symmetry,
the phase angles of these parameters will not be calculated in mean-field theory.
We, rather, perform a summation over all possible values of 
these angles. This procedure
ensures that the physical phase we obtain is nonmagnetic because it does not break
SU(2) symmetry.
\subsection{Effective Hamiltonian}
Using this new canal, we get the following approximate 
expression for the interacting term of (\ref{hamiltonian0}):
\begin{eqnarray}
{\bf S}_i\cdot{\bf s}_i\approx &&{1\over2}[S_i^- A_i^* + s_i^-Q_i^* + {\rm H.c.}
- 2{\rm Re}(Q_iA_i^*)],
\label{decouple}
\end{eqnarray}
where $Q_i=\langle S_i^-\rangle\equiv |Q_i|e^{i\phi_i}$
and $A_i=\langle s_i^-\rangle\equiv |A_i|e^{i\psi_i}$. Then,
the moduli $|Q_i|$ and $|A_i|$ are calculated in mean-field approximation, but
summation over the phase angles $\psi_i$ and $\phi_i$ are performed to guarantee 
that the continuous SU(2) symmetry remains unbroken.
The total Hamiltonian is 
averaged over $\phi_i$ and $\psi_i$, and
the summation over the phases $\phi_i$ and $\psi_i$ of the lowering spin
operators lead to
$\langle S_i^- \rangle_{\phi_i}=\int_0^{2\pi}{\rm d}\phi_i|Q_i|e^{i\phi_i}=0$ and
$\langle s_i^- \rangle_{\psi_i}=\int_0^{2\pi}{\rm d}\psi_i|A_i|e^{i\psi_i}=0$ where
$|Q_i|$ and $\phi_i$ on one hand, and $|A_i|$ and $\psi_i$ on the other hand
are considered to be independent variables.
The minimization of the average of the magnetic energy 
$J_K\langle{\bf S}_i{\bf s}_i\rangle=J_K|A_i||Q_i|\cos(\phi_i-\psi_i)$ imposes the
constraint $\phi_i-\psi_i=\pi$ on the angles. For this reason,
for example,
the sum over these phases in the last term of (\ref{decouple}), which involves
$A_i^*Q_i$, is 
$\int{d}(\phi_i-\psi_i-\pi)
|A_iQ_i|e^{i(\phi_i-\psi_i)}\delta(\phi_i-\psi_i-\pi)=-|A_iQ_i|$. The delta function
implements the constraint $\phi_i-\psi_i=\pi$.

Using the second quantization form for the 
itinerant spins, the Hamiltonian (\ref{hamiltonian0}) leads to the following
effective Hamiltonian where averaging 
over the phase angles $\psi_i$ and $\phi_i$ is done:
\begin{eqnarray}
{\cal H}=&&{J_K\over2}\sum_{i}\int_0^{2\pi}{{{\rm d}\psi_i}\over{2\pi}}
S_i^+|A_i|e^{i\psi_i} 
+ {\rm H.c.} \cr
&&+{J_K\over2}\sum_{i}\int_0^{2\pi}{{\rm d}\phi_i\over2\pi}\{{Q_i}
c^{\dag}_{i,\uparrow}c_{i,\downarrow}
+ Q_i^*c^{\dag}_{i,\downarrow}c_{i,\uparrow}\}\cr
&&-t\sum_{i,\sigma=\uparrow,\downarrow}c_{i,\sigma}^{\dag}c_{i+1,\sigma} 
+ {\rm H.c}\cr
&&-\sum_i \int{d}(\psi_i-\phi_i)
\times\cr
&&{J_K\over2}
\{
|A_iQ_i|e^{i[\phi_i-\psi_i]}\delta(\phi_i-\psi_i-\pi) + {\rm C.c.}\}.
\label{ham}
\end{eqnarray}
\subsection{Unifrom rotating configuration}
In the rest of this work, we focus our attention on the uniform configuration,
which is obtained for $Q_i=|Q|e^{i\phi}$ and $A_i=|A|e^{i\psi}$. $Q$, $A$, 
$\phi$ and $\psi$ are considered to be site independent but angles vary 
between $0$ and $2\pi$
while satisfying the constraint $\phi-\psi=\pi$. 
This state realized in this way is called the uniform rotating 
configuration (URC). In this state, the vector parameters 
$Q_i=\langle S_i^-\rangle$
and $A_i=\langle s_i^-\rangle$ point in opposite directions while rotating 
at the same rate. The local and total magnetizations are equal to zero.

In addition, we treat the up and down fermions as being different, and use the 
following canonical transformation:
\begin{eqnarray}
&&c_{k\uparrow}=e^{i\phi/2}(\rho_k + \sigma_k)/\sqrt{2}\cr
&&c_{k\downarrow}=e^{-i\phi/2}(\rho_k - \sigma_k)/\sqrt{2}
\label{transf}
\end{eqnarray}
in order to diagonalize the effective Hamiltonian (\ref{ham}) 
in the case of the URC. Here,
$c_k^{(\dag)}=\sum_ic_i^{(\dag)}e^{-ir_ik}/\sqrt{N}$ with $N$ being 
the number of lattice sites. Next,
we perform the following transformation to absorb the phase terms
in the transformations (\ref{transf}):
\begin{eqnarray}
c_{k,\uparrow}\to e^{i\phi/2}c_{k,\uparrow},\ \ 
c_{k,\downarrow}\to e^{-i\phi/2}c_{k,\downarrow}
\label{gauge}
\end{eqnarray}
which is equivalent to making a rotation by angle $\phi$ about the 
z-axis for the x- and y-components of the itinerant spin operator, with 
the matrix of rotation given by
\[
\left( 
\begin{array}{lll}
&\cos\phi\  &\sin\phi\\ 
-&\sin\phi\  &\cos\phi
\end{array}
\right).
\]
Due to the constraint $\phi-\psi=\pi$, rotating the itinerant-spins x and 
y components causes the rotation of the x and y components of the localized spins
by angle $\psi=\phi-\pi$ about th z axis. The matrix of rotation in this case
is given by:
\[
\left( 
\begin{array}{lll}
&\cos\psi\  &\sin\psi\\ 
-&\sin\psi\  &\cos\psi
\end{array}
\right).
\]
Because we sum over the angles $\phi$ and $\psi$ between $0$ and $2\pi$,
the spin components are continuously rotating, and the Hamiltonian may
be written in the rotating reference frame for which the x axis and y axis
coincide with the rotating x and y components of the itinerant spins.
\subsection{Mean field Hamiltonian in the rotating reference frame}
The fact that the effective Hamiltonian is invariant under the above rotations
by angles $\phi$ and $\psi$ is
consistent with the absence of SU(2) symmetry breaking.
Here we perform such rotations before
diagonalizing the simplified Hamiltonian obtained in the URC approximation.
This is equivalent to using the transformations
\begin{eqnarray}
&&c_{i,\uparrow}\to e^{i\phi/2}c_{i,\uparrow},\cr
&&c_{i,\downarrow}\to e^{-i\phi/2}c_{i,\downarrow}\cr
&&S^+_{i}\to e^{-i\psi}S^+_{i}
\end{eqnarray}
on both the itinerant and localized spins. The result is a much simpler expression
for the Hamiltonian:
\begin{eqnarray}
{\cal H}-\mu N=\sum_{k}&&\{E_\rho(k)\rho_{k}^{\dag}\rho_{k}
+ E_\sigma(k)\sigma_{k}^{\dag}\sigma_{k}\} \cr 
&& + AJ_K\sum_{i}S_i^x + J_K\sum_iAQ
\label{ham2}
\end{eqnarray}
where $A$ and $Q$ stand now for the magnitudes $|A|$ and $|Q|$ respectively.
Here, $E_{\rho,\sigma}=\epsilon({\bf k})-\mu \pm |Q|J_K/2$;
$\epsilon({\bf k})=-2t\cos k$ is the tight-binding spectrum, and $\mu$ is the
chemical potential.
Note that we should keep in mind that (\ref{ham2}) is obtained in a rotating
frame as $\phi$ and $\psi$ take values in the interval $[0,2\pi[$ subject to 
the constraint $\phi-\psi=\pi$. Therefore, it is inappropriate to interpret
this Hamiltonian as that of conduction electrons and localized spins
coupled to magnetic fields along the x-direction in a reference frame at rest.
Now, we understand that the up and down spins could be treated differently
as a consequence of the effective rotating magnetic 
fields $AJ_K$ for the localised spins, and $QJ_K$ for the itinerant spins.
Note also that the energy spectrum of the itinerant electrons 
splits into two bands under the effect of the Kondo
exchange interaction.
\subsection{Phase fluctuations}
Allowing for the summation over the phase angles means that we do not
minimize free energy with respect to these angles.
However, to satisfy the requirement that the magnetic energy is minimized
we constrained their 
difference to be $\pi$. This leads us to questioning whether this minimum 
is stable againt fluctuations about the value $\phi-\psi=\pi$.
To study the effect of these fluctuations, the following treatment is done. 
We replace the delta 
function in the last term of ${\cal H}$ in Eq. (\ref{ham}) by 
the broader Gaussian distribution 
$$
{1\over\epsilon\sqrt{\pi}}e^{-(\phi-\psi-\pi)^2/\epsilon^2}
$$
to allow for other
angles difference to contribute.
If $\epsilon\to0$ the Gaussian distribution reduces to the Dirac distribution.
The last term in (\ref{ham}) takes the form:
\begin{eqnarray}
&&\int{\rm d}(\phi-\psi) \cr
&&{J_K\over2}\sum_i \{
|AQ|e^{i[\phi-\psi]}{1\over\epsilon\sqrt{\pi}}
e^{-(\phi-\psi-\pi)^2/\epsilon^2} + {\rm C.c.}\}.
\nonumber
\end{eqnarray}
Integration 
over $\epsilon$ from $-\infty$ to $+\infty$ is undertaken, once the integration over 
$\phi-\psi$ is done, to guarantee 
that all possible fluctuations of the phase angles are embodied in 
the present approach. The result in the URC is
$$
-\alpha J_K\sum_i |AQ|
$$
which differs from the result $-J_K\sum_i |AQ|$, obtained without
fluctuations, by the factor $\alpha$. $\alpha$ s a number larger than 1 but 
smaller than  $2\sqrt\pi$ which is the value obtained when the integrations
are carried out to infinity in the integral on $\phi-\psi$.
This reduces the values of the mean-field parameters (as one would expect) 
by a factor $\alpha\sim1$ without
destroying the mean-field picture. Thus
this means that the mean-field solution $\phi-\psi=\pi$ 
is stable against phase fluctuations.
\section{Results}
\subsection{Parameters of the KLM model}
In the rest of this paper, we set $\alpha=1$, and seek some quantitative 
understanding of the present approach.
To calculate the parameters $Q$ and $A$, we use the self-consistent 
equations obtained
by minimizating the free energy:
\begin{eqnarray}
F=&&-{1\over N\beta}\sum_{k,\nu=\rho,\sigma}\ln\{1+e^{-{\beta}E_\nu(k))}\} \cr
&&-{1\over\beta}\ln\{2\cosh[\beta AJ_K/2]\} + J_KQA.
\end{eqnarray}
with respect to $A$ and $Q$. This
yields:
\begin{eqnarray}
Q=&&{1\over2}\tanh(\beta J_KA/2) \cr
A=&&{1\over2N}\sum_k\{f[E_\sigma(k)]-f[E_\rho(k)]\},
\label{mf}
\end{eqnarray}
where the summation $\sum_k$ runs over the Brillouin zone of the 
conduction electrons,
and $f(x)=1/(1+e^{\beta (x-\mu)})$ is the Fermi distribution factor.
$\beta = 1/k_BT$ with  $k_B$ is the Boltzmann constant.
$\mu\approx\epsilon_F$ with
$\epsilon_F$ being the Fermi energy of the conduction electrons.

As the paramagnetic phase occurs
for  $J_K< 4t$, we expand $f(E_\rho)-f(E_\sigma)$
to first order in $QJ_K/2$. We obtain:
\begin{eqnarray}
A=-{1\over2}{QJ_K}\sum_{k}
{\partial f(\epsilon({ k}))\over{\partial\epsilon({ k})}}
={1\over2}Q\chi J_K,
\label{mf1}
\end{eqnarray}
which leads to
\begin{eqnarray}
A=&&{1\over 4}\chi J_K \tanh(\beta AJ_K/2)
\label{mf2}
\end{eqnarray}
when (\ref{mf}) is used.
Here $\chi =-\sum_k
{\partial f(\epsilon(k))\over{\partial\epsilon(k)}}$ is the 
uniform susceptibility of the free conduction electrons in one dimension.

At zero temperature, the non-zero solution is given by $A=\chi J_K/4$ and 
$Q=1/2$; $\chi \approx{\cal D}(\epsilon_F)=1/2\pi v_F$ 
is the density of states of the free conduction electrons; $v_F=2t\sin k_F$ being 
their Fermi velocity.
Slightly above zero temperature, we get for $k_BT\ll{\cal D}(\epsilon_F) J_K^2/4$
\begin{eqnarray}
&&A \simeq {1\over4}\chi J_K(1-2e^{-\beta\chi J_K^2/4}) \cr
&&Q \simeq {1\over2} - e^{-\beta\chi J_K^2/4}
\nonumber
\end{eqnarray}
where again $\chi\approx{\cal D}(\epsilon_F)$ is evaluated at $T=0$ because
$\chi(T)$ does not deviate much from its zero-temperature value for 
temperature well below the Fermi temperature of the free conduction electrons.
In the rest of this paper we will consider this value for $\chi$.
\subsection{Crossover temperature}
The parameters $A$ and $Q$ decrease as temperature increases, and vanish
at a temperature $T_{cf}$. Indeed, expanding the tangent 
hyperbolic to third order 
in the vicinity of this critical temperature in Eq.\ (\ref{mf2}) and 
using Eq. (\ref{mf2}), we find:
\begin{eqnarray}
A \simeq A_0\{1-T/T_{cf}\}^{1/2},\ 
Q \simeq Q_0\{1-T/T_{cf}\}^{1/2},
\nonumber
\end{eqnarray}
with $A_0=2\sqrt{3}k_BT_{cf}/J_K$, and $Q_0=\sqrt{3}/2$,
and the temperature
\begin{equation}
T_{cf}=\chi J_K^2/8k_B.
\label{tx}
\end{equation}
Below $T_{cf}$,
the Kondo lattice system deviates
from the system of independent localized spins and conduction electrons.
As might be expected $T_{cf}$ is proportional
to $\chi J_K^2$, a result that is reminiscent of the RKKY interaction which
dominates over Kondo screening, although not leading to magnetic order.
Above $T_{cf}$, the magnitude of the magnitization is zero. Below, $T_{cf}$,
the magnitude becomes finite, but averaging about its phase angles gives zero 
magnitization. It is for this reason that this change in behavior is not a 
true phase transition. The present appoach leads to a sharp change of regime at
$T_{cf}$. We do not exclude that 
if the change in the regime is rather a smooth crossover,
the crossover temperature will be given by $T_x\sim T_{cf}$.
$T_{cf}$ is called the crossover temperature.
Shibata and Tsuntsugu\cite{shibata2} reported the 
existence of a crossover temperature, but said that it is determined by the 
velocity of the excitations of the itinerant spins at zero temperarture.
\subsection{Ground-state energy and the phase diagram}
The idea behind calculating the 
ground-state energy is to find
how the correction (due to $J_K$) to the ground-state energy 
of the conduction electrons behaves. This will indicate the
type of correlations that dominant.
This correction is found to be given by the equation:
\begin{eqnarray}
\Delta E_{GS}
\approx -{J_K^2\over8}{\cal D}(\epsilon_F),\ \ \ 
(|\Delta E_{GS}|/k_BT_{cf}\approx1).
\end{eqnarray}
for $J_K\ll 4t$.
Interestingly, this correction is of order $J_K^2/t$, and is 
again reminiscent of
the RKKY interaction between localized spins. 
As $J_K/t$ increases (but keeping $J_K/t<4$),
we found that the correction to the ground-state energy 
deviates from a $J_K^2/t$ law.
Estimating the points where the change takes place for different values of $J_K$,
we determined the line boundary, $J_{K,c}(n)/t$ 
versus the conduction electrons density $n$, separating the paramagnetic 
phase from the ferromagnetic 
phase at strong coupling $J_K/t$. This line is found to be very well fitted by:
\begin{equation}
J_{K,c}(n)/t\approx2\pi n\sin(\pi n).
\label{fm}
\end{equation}
This is consistent with the fact that 
the weak coupling regime is equivalent to 
$J_K/t<\epsilon_F/t$. 
Linearizing the energy spectrum of the conduction electrons around
$k_F$, one finds the linear spectrum $\epsilon(k) -\epsilon_F=v_F(k-k_F)$,
where $v_F$ is the Fermi velocity.
For a particle with momentum $k$ ($\hbar=1$) and velocity $v_F$,
the energy is given by $v_Fk_F$. Thus, within the linear approximation, 
it is  natural to consider an effective
Fermi energy given by $\epsilon_F=v_Fk_F=2tk_F\sin k_F$.
This yields $J_K<v_Fk_F=2t\pi n\sin(\pi n)$, which leads to
(\ref{fm}).

Eq. (\ref{fm}) is in very good agreement with the available 
exact numerical data, Fig.\ 1.
For $n=0.35$, Eq. (\ref{fm}) yields $J_{K,c}\approx1.96$, 
which is very close to the density-matrix-renormalization-group
exact result, $2$\cite{moukouri2}. The agreement 
is however less accurate
for $n=1/6$ where we obtained $J_{K,c}\approx0.5$ compared to 
the result of Ref.\cite{troyer}, namely $J_{K,c}\approx1$, 
obtained using quantum 
Monte Carlo simulation. The reason for this discordance 
may however be attributed to the fact that Monte Carlo
simulations are difficult to implement at very low temperatures, and to the fact 
that the present theory is mean-field like.
Using the bosonization 
technique\cite{honner} Honner and Gulacsi, reported
$J_{K,c}(n)\approx2.5\sin(\pi n)$. But as we notice on figure 1,
this does not capture the general trend, while Eq.\ (\ref{fm}) does it fairly
well. Honner and Gulacsi noted that a dependence as that I report here
is also possible.
\begin{figure}
\centerline{\psfig{figure=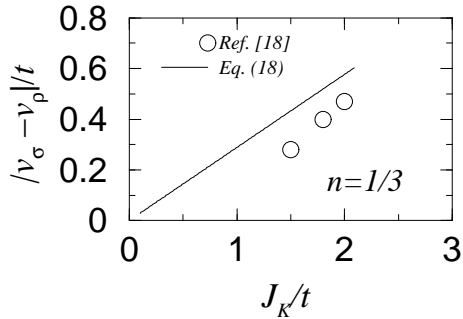,height=4.5cm,angle=0}}
\caption{The difference between the spin and charge velocities
of the conduction electrons is displayed in terms of $J_K/t$.}
\label{DELTAVF}
\end{figure}
\subsection{Excitation velocities}
For densities below the half-filled band, and in the presence of 
the Kondo coupling, we have found that 
the conduction electrons as charge and spin entities 
still form a gapless phase because the term $\pm |Q|J_{K}/2$
in $E_{\rho,\sigma}$ does not open a gap in the energy spectrum. 
The elementary excitations are described by the operators $\rho_k$
and $\sigma_k$. Their excitation velocities are given by 
$v_{\rho,\sigma}=2t\sin k_{\rho,\sigma}$ where
$k_{\rho,\sigma}=n\pi\mp J_K/8t\sin(n\pi)$ for small $J_K$. 
$k_{\rho}$ and $k_{\sigma}$ are determined by 
the conditions $E_{\rho}=\epsilon_F$
and $E_{\sigma}=\epsilon_F$ with $\epsilon_F=-2t\cos(k_F)$.
Thus, the $\rho$- and $\sigma$-velocities are given 
by $v_{\rho}\simeq v_F-J_K\cot(n\pi)/4$
and $v_{\sigma}\simeq v_F+J_K\cot(n\pi)/4$ for 
$J_K<J_{K,c}(n)$. This leads to
\begin{eqnarray}
\Delta v=v_{\sigma}-v_{\rho}\simeq {J_K\over 2}\cot(n\pi).
\label{velocities}
\end{eqnarray}
It is not clear how we could relate these velocities
to the spin and charge velocities of the conduction electrons
obtained within the bosonization approach. 
In Fig.\ 2, we draw $\Delta v$ in terms of $J_K$ for $n=1/3$, and compare it with
Shibata {\it et al.}'s\cite{shibata} density-matrix-renormalization-group 
results. Note that this comparison should be considered with caution
as our results for $v_\rho$ and $v_\sigma$ disagree with theirs. Only the 
difference compares well. We should also emphasize here that we do not 
pretend by any means
that the present mean-field theory is equivalent to the bosonization technique
which is used to obtain the spin and charge velocities when 
spin-charge separation takes place in a system of interacting fermions.
\subsection{Thermodynamic functions}
Finally, we would like to briefly analyze the entropy $S=-\partial F/\partial T$ 
and heat capacity $C= T\partial S/\partial T$.
In Fig.\ 3, we display $C$, for $n=0.35$ and $J_K/t=1.6$
where $T_{cf}=0.057t/k_B$ agrees very well with the crossover temperature 
predicted by Shibata and Tsunetsugu  in their numerical work \cite{shibata2}.
The entropy $S$ is also reported on the same figure. It behaves as predicted
in Ref.\cite{shibata2}.
For $T/T_{cf}\to0$, the entropy $S\simeq1.24T$ yields a slope in
very good agreement with the Tomonaga-Luttinger result\cite{takahashi}
$S=\pi T(v_\sigma^{-1} + v_\rho^{-1})/3\simeq1.20T$.
A sharp maximum is found in $C$ at $T_{cf}$ as a consequnece of the change in
the regime from hight temperatute to temperatures below $T_{cf}$.
The linear behavior of entropy in terms of temperature is consistent with the gapless
quasi-particle excitations, and is an indication of
the {\it metallic} character of the paramagnetic state. 
\begin{figure}
\centerline{\psfig{figure=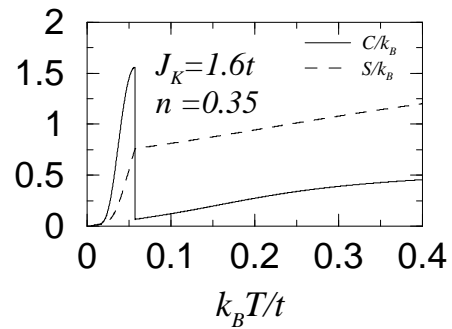,height=4.5cm,angle=0}}
\caption{The specific heat and entropy
are drawn vs $T/t$ for $J_K=1.6t$ and $n=0.35$.
Comparison is made to the results of Ref. 20.}
\label{CvS}
\end{figure}
\section{conclusions}
In summary, the paramagnetic phase of the one-dimensional Kondo lattice model
is investigated using a new mean-field approach. This 
approach is valid away from half-filling where Kondo screening is
negligible. We predict that 
this phase is described by a new nonmagnetic-quantum state 
characterized by two rotating 
order parameters with nonzero magnitudes below a crossover temperature.
We calculated the ground-state line boundary in the density-coupling phase diagram,
the elementary-excitations velocities of the conduction electrons,
heat capacity, and entropy.
The crossover temperature separating the low-temperature paramagnetic quantum 
phase from the normal high-temperature phase is evaluated as a function of the 
Kondo coupling constant and the density of states of the conduction electrons. 
Overall, our results agree very well with many of 
the available numerical data well below 
half-filling, and are compatible with the Luttinger-liquid picture as put forward
by several authors, (see Ref. \cite{tsunetsugu1} and references therein).
Note that it is not yet clear how we could interpret the results of our 
approach in terms of the Luttinger-liquid description.
For densities close to half-filling, residual effects due to Kondo screening
have to be taken into account. It is then natural that, quantitatively, our approach
is less accurate in this limit. This is what we come to face when we compare
Shibata and Tsunetsugu's results\cite{shibata2} with ours
for the entropy $S$ for example.
The results start to agree well only for conduction-electrons 
densities below 0.35.
Finally, the RKKY oscillations constitute an issue 
that needs to be addressed within
the present approach. One possible avenue
is to allow for the parameters $A$ $Q$ to be $k$-dependent.
However, even with these two limitations, the present approach is very promising
because it is very simple to use, and the calculations can, to a very large extent,
be done analytically. And most importantly, it leads to many 
very satisfactory physical results.

The author would like to thank Prof. P. Fulde for his comments
on the manuscript.


\begin{references}
%
\bibitem{tsunetsugu1} H. Tsunetsugu, M. Sigrist, and K. Ueda, 
Rev. Mod. Phys. {\bf 69}, 809 (1997), and references therein.
%
%
\bibitem{doniach} S. Doniach, Physica {\bf 91B}, 231 (1977).
%
\bibitem{hirsh} J. E. Hirsh, Phys. Rev. B {\bf 30}, 5383 (1984).
%
\bibitem{noziere} P. Nozi\`ere, Ann. Phys. (Paris) {\bf 10}, 19 (1985).
%
\bibitem{sigrist} M. Sigrist, H. Tsunetsugu, and K. Ueda, Phys. Rev. Lett. 
{\bf 67}, 2211 (1991).
%
\bibitem{fazekas} P. Fazekas and E. Muller-Hartmann, Z. Phys. B {\bf 85}, 
285 (1991).
%
\bibitem{fye} R. M. Fye and D. J. Scalapino, Phys. Rev. B {\bf 44}, 7486 (1991).
%
\bibitem{sigrist2} M. Sigrist, H. Tsunetsugu, K. Ueda, and T. M. Rice, 
Phys. Rev. B {\bf 46}, 13838 (1992).
%
\bibitem{rise} P. S. Riseborough, Phys. Rev. B {\bf 45}, 13984 (1992).
%
\bibitem{troyer} M. Troyer and D. Wurtz, Phys. Rev. B {\bf 47}, 2886 (1993).
%
\bibitem{tsunetsugu} H. Tsunetsugu, M. Sigrist, and K. Ueda, Phys. Rev. B 
{\bf 47}, 8345 (1993);
%
\bibitem{yu} C. C. Yu and S. R. White, Phys. Rev. Lett. {\bf 71}, 3866 (1993).
%
\bibitem{moukouri2} S. Moukouri and L. G. Caron, Phys. Rev. B {\bf 52}, 
R15723 (1995).
%
\bibitem{moukouri} S. Moukouri, L. Chen, and L. G. Caron, Phys. Rev. B 
{\bf 53}, R488 (1996).
%
\bibitem{honner} G. Honner and M. Gul\'acsi, Phys. Rev. Lett. {\bf 78}, 
2180 (1997) and references therein.
%
\bibitem{caprara} S. Caprara and A. Rosengren, Euro. Phys. Lett., {\bf 39}, 55 (1997).
%
\bibitem{sikkema} A. E. Sikkema, I. Affleck, and S. R. White, Phys. Rev. Lett. 
{\bf 79}, 929 (1997).
%
\bibitem{shibata} N. Shibata, A. Tsvelik, and K. Ueda, Phys. Rev. B {\bf 56}, 330
(1997).
%
\bibitem{haldane} F. D. M. Haldane, J. Phys. C {\bf 14}, 2585 (1981).
%
\bibitem{shibata2} N. Shibata and H. Tsunetsugu, preprint, cond-mat[9812191]
%
\bibitem{takahashi}M. Takahashi, Prog. Theor. Phys. {\bf 47}, 69 (1972);
{\bf 52}, 103 (1974).And T. Usuki, N. Kawakami, and A. Okiji, Phys. Lett. {\bf 135A},
476 (1989).

\end{references}
\end{document}